# Guided search for desired functional responses via Bayesian optimization of generative model: hysteresis loop shape engineering in ferroelectrics

Sergei V. Kalinin,[1] Maxim Ziatdinov[1,2], and Rama K. Vasudevan[1]

[1]The Center for Nanophase Materials Sciences, Oak Ridge National Laboratory, Oak Ridge, TN 37831

[2]The Computational Sciences and Engineering Division, Oak Ridge National Laboratory, Oak Ridge, TN 37831

Advances in theoretical modelling across multiple disciplines have yielded generative models capable of high veracity in predicting macroscopic functional responses of materials emerging as a result of complex non-local interactions. Correspondingly, of interest is the inverse problem of finding the model parameter that will yield desired macroscopic responses, such as stress-strain curves, ferroelectric hysteresis loops, etc. Here we suggest and implement a Gaussian Process based methods that allows to effectively sample the degenerate parameter space of a complex non-local model to output regions of parameter space which yield desired functionalities. We discuss the specific adaptation of the acquisition function and sampling function to make the process efficient and balance the efficient exploration of parameter space for multiple possible minima and exploitation to densely sample the regions of interest where target behaviors are optimized. This approach is illustrated via the hysteresis loop engineering in ferroelectric materials, but can be adapted to other functionalities and generative models. The code is open sourced and available at [github.com/ramav87/Ferrosim].

[1] Sergei2@ornl.gov

**Introduction**

Since the discovery of BaTiO$_3$ solid solutions in the late 1940s, ferroelectric materials have become one of the central topics in condensed matter physics and materials science alike.[1-4] Devices such as SONARs and later piezoelectric and optoelectronic ceramics have attracted the strong interest from application community, while the unique physics of this materials have stimulated intensive fundamental research and theoretical studies of these material. Initially, fundamental research was focused on the properties of single crystals of simple ferroelectrics,[5] whereas many applications were preponderantly based on the ceramic materials.[6]

With the development of the field, the attention of scientific community has been shifted towards the materials systems such as disordered ferroelectrics,[7-9] morphotropic phase boundary materials,[10,11] and ferroelectric relaxors.[12-20] These materials generally offer giant dielectric and electromechanical responses, both of interest to applications and necessitating understanding the fundamental origins of these phenomena. After multiple studies, it was established that a common feature of these systems is the presence of nanoscale polarization inhomogeneities[12,21-24] emerging due to the interplay between polarization instabilities, mesoscale elastic and electric depolarization fields, and structural defects. Correspondingly, the materials response to even small perturbations can result in strong changes in the inhomogeneous ground state of the system, giving rise to a rich spectra of phenomena such as broad relaxation time distributions, strong frequency dispersion of responses, giant responses, smeared phase transitions, etc. These behaviors are not limited to disordered ferroelectrics and similar behaviors are observed in other materials including phase separated manganites, spin and cluster glasses, etc.[25-30]

However, these developments have further brought forth the challenge of understanding these materials. In classical ferroelectrics that are homogeneous on the length scales between unit cell and domain size, the functionalities can be readily described using mesoscopic Ginzburg Landau Devonshire (GLD) type theories with the parameters readily available from macroscopic measurements.[5,31-36] More recently, these can also be obtained from the density functional theory (DFT) models. The gradient and interfacial terms necessary for any phase-field type model are often postulated, or can be derived from atomistic imaging.[37,38] Jointly, this combination of DFT and GLD allows comprehensive description of ferroelectric materials and microstructures on all length scales. With these, materials design and optimization on the unit cell level can be approached via a theory guided search through the chemical space of existing materials.

The situation is fundamentally different for the disordered ferroelectrics. Here, the functionalities emerge as a result of collective phenomena on multiple length scales, including processes such as polarization rotations and ferroelastic domain wall motion. Often, these mechanisms are defined only via macroscopic descriptors, whereas associated local mechanisms are virtually unknown. The existence of order parameter that can describe these behaviors is heavily debated,[17,39-41] as is the case for other disordered systems. Finally, the applicability of DFT type models for the description of the stochastic mesoscopic systems is naturally limited.

These considerations have stimulated the intensive development of the mesoscopic lattice type models, where the functionality of the material is represented via collection of spin-like variable on the periodic lattice. The properties (scalar, vectors) and corresponding dynamics are introduced, and time dynamics of the system for various field, temperature, and time histories is explored using classical Monte-Carlo methods.[10,42,43] While these models provided the certain insight into the evolution and properties of the ferroelectric and relaxor state, the nature of the relaxor states, appropriate reduced descriptors, and especially pathways to tailor macroscopic functionalities via microstructure have not been explored.

Here, we explore the targeted optimization of the relaxor functionalities via microstructural engineering. As the descriptors for relaxor behavior, we choose the strength and concentration of the defects and the parameters of the embedding GLD functional. In this description, the global polarization behavior emerges as a collective response of polarization fields interacting with the defects. We show that the Gaussian Process optimization can be used to explore the phase space of this system and discover the functionalities of interest. This approach allows to both explore the general parameter space and localize target areas of interest (of which there can be more than one), as demonstrated here for the macroscopic observable of microscopic model. We note that while here this analysis is performed for random defect distribution, this analysis can be performed for microstructural models based on interacting defects or sampled from generative models derived from experimental observations, for example from direct imaging studies, or from atom probe tomography, pair distribution functions from scattering, etc.

**Results and Discussion**

As a model of ferroelectric material, we chose the lattice model introduced by Ricinschi et al.[44] Here, the polarization at each lattice site is represented as a continuous scalar variable, $p$, as shown in Fig. 1 (a). The local energy as a function of this variable follow the standard GLD form, $F_{loc} = (\alpha/2)p^2 + (\beta/4)p^4 - E_{loc}p$, where $\alpha = \alpha_0(T - T_C)$ and $\beta$ are GLD coefficients and $T_C$ is Curie temperature. The term $-E_{loc}p$ describes the usual coupling between polarization and the local electric field.

The collective effects in this model are introduced via the gradient like interactions,

$$F = \sum_{i,j}^{N}\left(F_{loc} + K\sum_{k,l}(p_{i,j} - p_{i+k,j+l})^2\right), \tag{1}$$

where $i, j = 1, .. N$ are the lattice sites, $K$ is the coupling constant, and the sum over $k, l$ is taken over chosen neighborhood for each site.

The disorder in the model can be introduced in multiple ways, e.g. via the position dependence of coupling constant $K$ (bond disorder), or presence of the local electric fields $E_d(i,j)$ (field disorder). The depolarization field effects are introduced via spatially uniform depolarization field $E_{dep} = -\alpha_{dep}\langle P \rangle$, where $\alpha_{dep}$ is the depolarization factor and $\langle P \rangle$ is averaged polarization. In realistic systems, $\alpha_{dep}$ is determined by the geometry of the system and efficiency of screening; here we consider it to be a control parameter. The local field acting on each lattice site is then the sum of external, depolarization, and disorder fields, $E_{loc} = E_{ext} + E_{dep} + E_d(i,j)$. Note that the separation of the acting field into the local on-site component and averaged off-site component is the simplest possible mechanism accounting for the physics of the problem.

Finally, the dynamics of the local spins is given by the classical Landau-Khalatnikov equation, $\gamma dp_{ij}/dt = -\partial F/\partial p_{ij}$. The output of the models can be calculated in the usual fashion and can represent the average polarization $\langle P \rangle$, its evolution as a function of external control stimuli, or some descriptors of microscopic states.

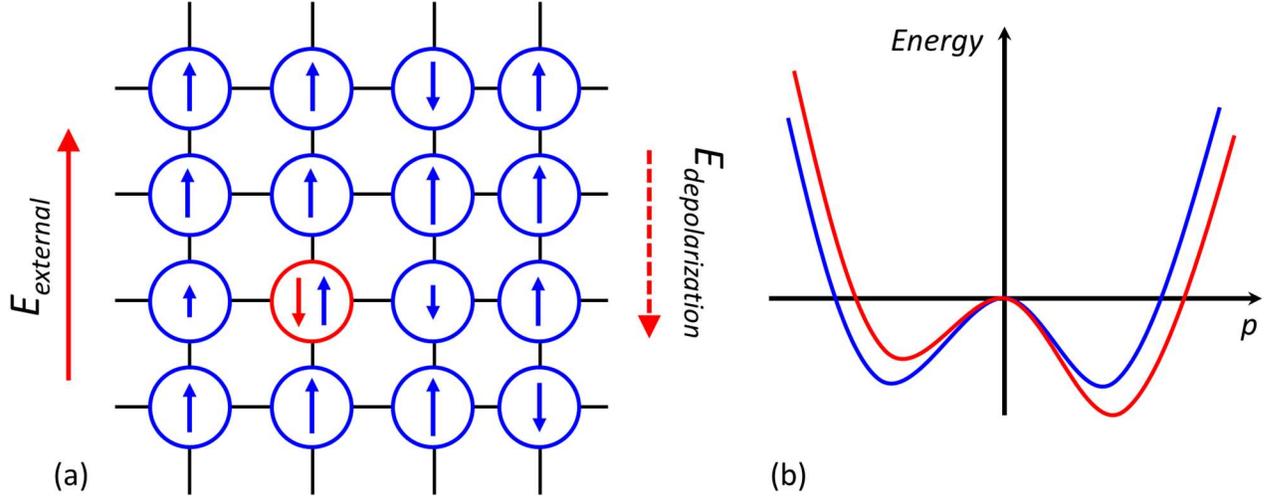

**Figure 1.** (a) Schematic representation of the lattice GLD model and (b) free energy of each spin with (red) and without (blue) random field disorder.

Here the model is realized on a square lattice of size $N*N$, with the starting polarization at each site equal to the nominal remnant state of the system, and given by the solution to (1) in the absence of field, i.e. $p_r = -\sqrt{-\alpha/\beta}$. An applied electric field, in this case a simple sine wave with prescribed frequency and amplitude is used. The simulation begins by selecting a lattice site, and then computing the gradient at time $t_n$:

$$\frac{dp_{ij}}{dt} = -\gamma^{-1}\left(\beta p_{ij}^3 + \alpha p_{ij} + K \sum_{k,l}(p_{ij} - p_{kl}) - E_{loc}\right) \qquad (2)$$

The polarization at the chosen site is updated in the direction of the gradient, i.e. $p_{ij}(t_{n+1}) = p_{ij}(t_n) + \Delta t * \frac{dp_{ij}}{dt}$. Once the polarization is updated across all lattice sites for $t_n$, the simulation progresses to the next time step where the same loop resumes. The output of the simulation is the polarization evolution of the system, discretized in space and time, i.e. a matrix of size $(t,n,n)$ where $t$ is the number of time steps of the simulation, and $n$ is the simulation size (assuming a square lattice of size $n$ x $n$). It should be noted that this lattice model represents the intermediate case between the fixed spin lattice models such as Ising or Heisenberg models, and the classical phase-field approach. Compared to the Ising model, the spin variable now is continuous and follows the classical Khalatnikov equation for order parameter. At the same time, the depolarization fields are assumed uniform, unlike the PFM models. The latter approximations allows avoiding the use of semi-implicit spectral methods necessary to recalculate the depolarization field distributions at each step, severely reducing the calculation time. That said, the approach developed here is universal and can be applied for any generative model, as limited by available computational capabilities.

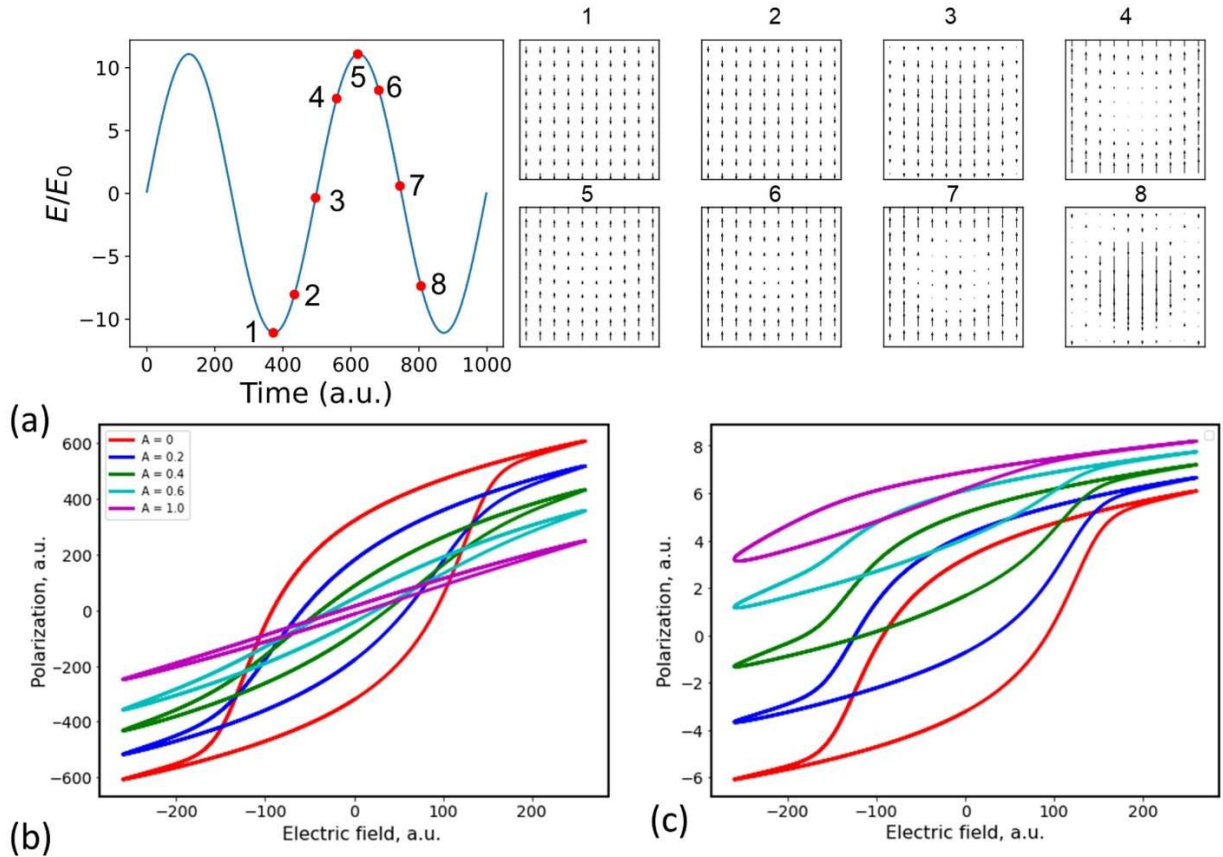

**Figure 2.** (a) Sequence of the polarization states of the 10x10 lattice during the sinusoidal sweep of the electric fields. (b) Evolution of the hysteresis loops as a function of the depolarization factor. (c) Evolution of the hysteresis loops as a function of the defect concentration.

To illustrate the behavior of the model, shown in Fig. 2(a) is the evolution of the microscopic states of the system during the polarization switching induced by external electric field. Here, the domain nucleation, formation of domain walls, and interaction of domain walls with the random field defects are clearly visible.

The evolution of the average polarization as a function of the field is show in Fig. 2 (b) for different depolarization factors. Here, classical well-saturated hysteresis loops for zero depolarization are obtained. On increasing the depolarization factors, the loops become more constrained and ultimately collapse to a straight line, as expected for a ferroelectric material. Similarly, evolution of the hysteresis loop as a function of the random field strength is shown in Fig. 2 (c), exhibiting successful narrowing of the hysteresis loop and pinning to the preexisting defects. Similarly, effects of other control variables including the concentration of the defects, strength of local coupling, or GLD coefficients can be explored.

With this model in hand, we aim to explore which combinations of parameters can be used to yield the desired functionalities or microstructures. For ferroelectric material, desired functionality can include certain characteristics of the hysteresis loop. These can include classical descriptors such as hysteresis loop area, remanent polarization, imprint, or asymmetry. Alternatively, ad hoc descriptors can be introduced to describe the deviation of the loop from square, maximum slope, etc. as required by specific applications. Similarly, descriptors can be constructed for the more complex field histories, for example specific Preisach density characteristics. Finally, descriptors can be constructed based on spatial variability of polarization field and its field history dependence, e.g. variability of polarization field for specific voltage and field history, critical bias for domain formation, etc.

It is important to note that the system behavior and associated descriptors can be sensitive to the details of the microstructure, i.e. the exact positions of the defects in this case. It is generally expected that macroscopic descriptors such as loop shape will be robust with respect to exact microscopic defect configurations and be determined by the average concentrations and defect strength, whereas specific polarization field distributions are not. These considerations will be important for future discussion of materials design.

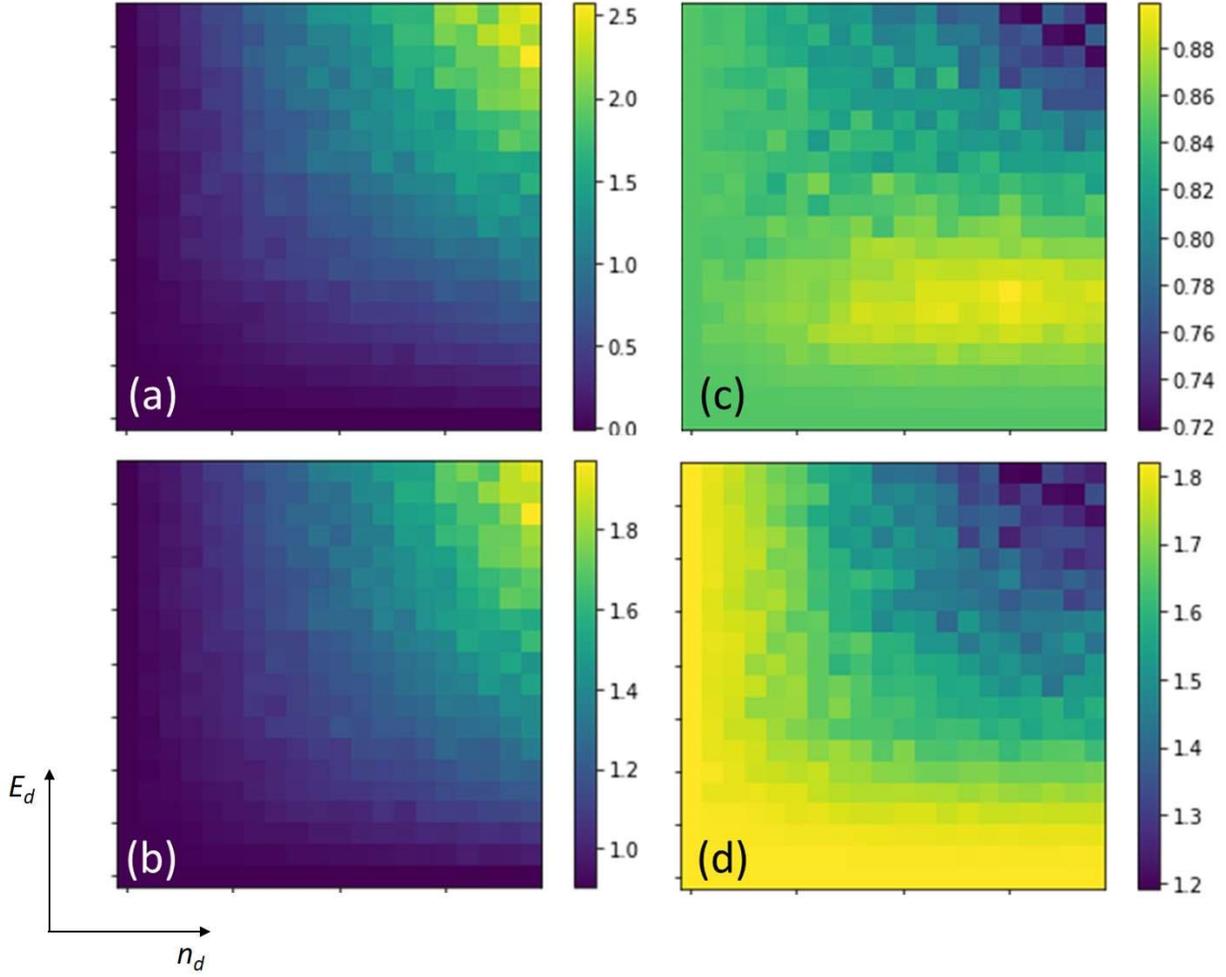

**Figure 3.** Evolution of the hysteresis loop parameters including (a) total polarization (integral of the upper branch of the loop), (b) remanent polarization on the upper branch of the loop, (c) area under the loop, and (d) switchable polarization with the number of defects ($n_d$, horizontal axis) and random field strength ($E_d$, vertical axis), both in the (0,100) interval.

In this setting, the material design for specific applications becomes the inverse problem of finding the model parameters that maximize the required functionality, i.e. specific hysteresis loop or robust microscopic descriptor. The direct approach for solving such problem can be based on direct grid search in the parameter space, as illustrated in Fig. 3 for dependence of hysteresis loop parameters on the defect strength and concentration. From the graphs, it is obvious that here with a good degree of approximation the required properties will be uniform along the lines $n_d E_d$ = *const*, as can be surmised from the fact that in this case domain formation is impeded and hence total defect field emerges as a control variable. This observation, along with the available control variables such as dopant solubility and tendency for clustering, can be used as a basis for materials selection.

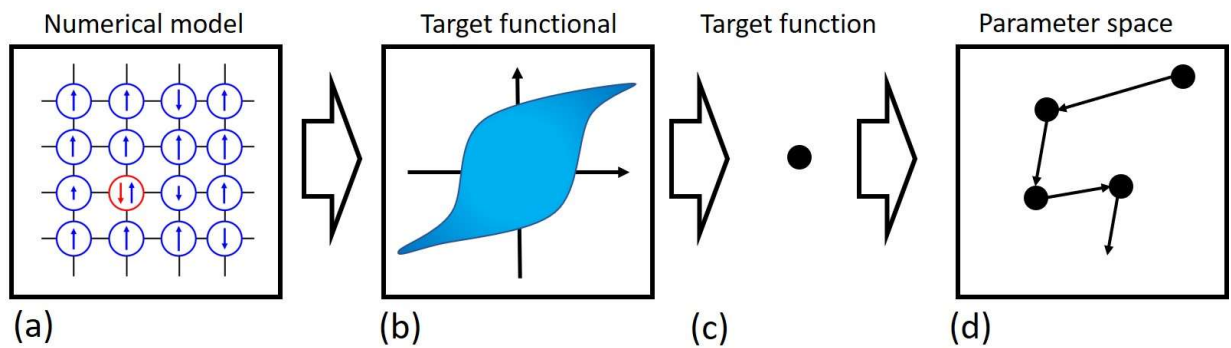

**Figure 4.** The Bayesian optimization is used to discover the region of the parameter space of the non-local model that gives rise to the target functional behavior. Here, (a) the numerical model is evaluated to give rise to (b) functionality of interest. Here, we choose the lattice ferroelectric model and ferroelectric hysteresis loop, but these can be arbitrary. (c) Based on the target functional, we evaluate a single scalar parameter that represents the "goodness" of the system for applications. (d) Bayesian optimization is used to efficiently explore parameter space for the model balancing the exploration and exploitation, i.e. targeting the optimum value of the target function.

However, exploration of the parameter space of the system targeting desired functionalities is a highly non-trivial problem. The dimensionality of parameter space even for the simplified model used here precludes simple grid search approach. At the same time, the methods based on the gradient descent tend to behave poorly when the basins of attraction have complex shape and for the case of the behaviors in Fig. 3 will yield a single combination of $n_d$ and $E_d$, while the presence of other solutions tend to remain undiscovered. Finally, if the parameter space contains multiple optima corresponding to disparate degenerate low-dimensional manifolds in high dimensional parameter spaces, many of these will remain undiscovered.

Here, we explore the use of Bayesian optimization (BO) for the exploration of these complex parameter spaces to search for target functionalities (Fig. 4). The BO is underpinned by the Gaussian Process (GP) regression methods. Below, we briefly introduce the salient aspects of GP and describe how to build the BO search on top of it.

The GP refers to a general problem of learning a function, $f$, from given set of observations $D = \{(x_1, y_1), \ldots, (x_N, y_N)\}$. It is assumed that observations are noise-corrupted values of the function, $y = f(x) + \varepsilon$, where $\varepsilon$ is Gaussian observation noise, whereas the arguments are known exactly. The learning is performed via Bayesian inference in a function space. The key assumption in this analysis is that the function $f$ has a prior distribution $f \sim \mathcal{GP}(0, K_f(x, x'))$, where $K_f$ is a covariance function (kernel).[45] The kernel function defines the strength and functional form of correlations between the values of the function across the parameter space. The functional form of the kernel is postulated in the beginning of the regression process. In the cases when physical guidance is absent, the typical choice can be the Gaussian (or Radial Basis Function, RBF) kernel.

During the GP regression, the expected value of the function, corresponding uncertainties, and kernel hyperparameters are optimized simultaneously. The output of the GP process is then the predicted data set and uncertainty maps representing the quality of prediction. Additionally, kernel parameters can yield the information on the physics of the system.

The GP can be further used for the targeted exploration of the parameter space. Here, the process starts from a small number of seed points that serve to define initial guesses on kernel hyperparameter, function values, and uncertainties. With these, the subsequent points in the parameter space can be selected. In a purely exploratory strategies, the points are selected such as to minimize the uncertainty in the system. In the exploration-exploitation strategies, the acquisition function that includes both the expectation values of the parameter and uncertainty is formed and the process is driven to minimize this acquisition function. The choice of the acquisition function for each specific problem requires tuning to avoid the process to be stuck in a certain region of the parameter space. Thus, GP allows to both explore the general parameter space, and localize target areas of interest (of which there can be more than one).

Here we implemented the BO with GP based on the GPim package.[46] We used the RBF or Matern kernels, defined as

$$k_{RBF}(x_1, x_2) = \sigma^2 \exp\left(-0.5 \times \frac{|x_1 - x_2|^2}{l^2}\right) \qquad (2)$$

and

$$k_{Matern}(x_1, x_2) = \sigma^2 \exp\left(-\sqrt{5} \times \frac{|x_1 - x_2|}{l}\right)\left(1 + \sqrt{5} \times \frac{|x_1 - x_2|}{l} + \frac{5}{3} \times \frac{|x_1 - x_2|^2}{l^2}\right), \qquad (3)$$

where $l$ and $\sigma^2$ are kernel length scale and variance, respectively, which are learned from the data by maximizing the log-marginal likelihood.

Here, we explored the effect of the defect concentration, defect strength, and depolarization field on the hysteresis loop shape. The 2D or 3D parameter space is defined as dense uniform grid of possible parameter values. Note that the function values are not evaluated initially and the grid is used only as discretization of the parameter space. We subsequently implement the Bayesian optimization strategy exploring the parameter space based on the maximum uncertainty (exploration), searching for target function values (exploitation), or both.

To avoid the exploration of only narrow region of parameter space once the target function value is pursued in exploitation strategy, we have introduced the approach where the next measurement point cannot be chosen closer then given distance $l$ to the previous point. The distance can be set as equal to the kernel length scale or defined *ad hoc*. Subsequent experimentation has suggested that optimal results can be obtained if the short-term memory is introduced, namely that location at time *t* is chosen no closer than $l$ to the location at step *t*-1, $al$ at step *t*-2, $a^2l$ at step *t*-3 and so on. Here, we have chosen the length of the memory to be equal to 10 and memory decay coefficient $a = 0.8$; however other values are possible.

Finally, to combine the exploration of the broad parameter space and maximization of required behavior, we have introduced combined strategy where the acquisition function was switched every ten steps from purely exploratory to the exploitation. In the exploratory regime, the points in the parameter space are chosen based on the maximum uncertainty. In the exploitation step, the algorithm maximized the target function, defined here as

$$Lc = exp(-(mean - target)^2), \qquad (4)$$

where mean is the expected value of the function from GP regression, and target is the target value of specific descriptor. The effect of the target function is to focus exploration in the locations where the expected value of the function is closest to target.

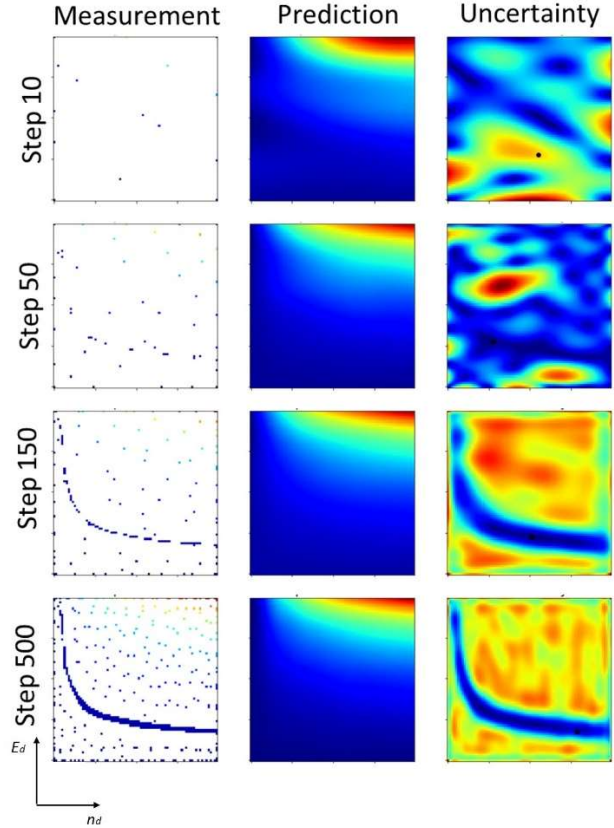

**Figure 5.** GP-based optimization of the hysteresis loop asymmetry targeting the value 0.5.

To illustrate the Bayesian optimization of the hysteresis loop shape, shown in Fig. 5 is the evolution of the search process in the two-dimensional parameter space for $n_d$ in the range (0,80) and $E_d$ in the range (0,80). This range covers the complete evolution of the hysteresis loop from completely open for $(n_d, E_d) = (0,0)$ to collapsed at (80,80). As a target function, we have somewhat arbitrarily chosen the loop offset being 0.5, corresponding to the remanent polarization on the lower branch of the loop being zero.

At the initial stages of search (10 steps), the measurement points are randomly distributed in the parameter space and the corresponding uncertainty map determined by the convolution of measurement points and kernel function shows large variability. Upon sequential evolution (step 50), the algorithm identifies the manifold on the parameter space where the function adopts the target value. From this moment, the exploitation step preponderantly focuses on the points in the vicinity of this manifold, whereas the exploration steps continue sampling the parameter space (steps 150 and 500). The kernel hyperparameters in this process converge relatively rapidly and remain stable during the optimization, reflecting the monotonic character of response function.

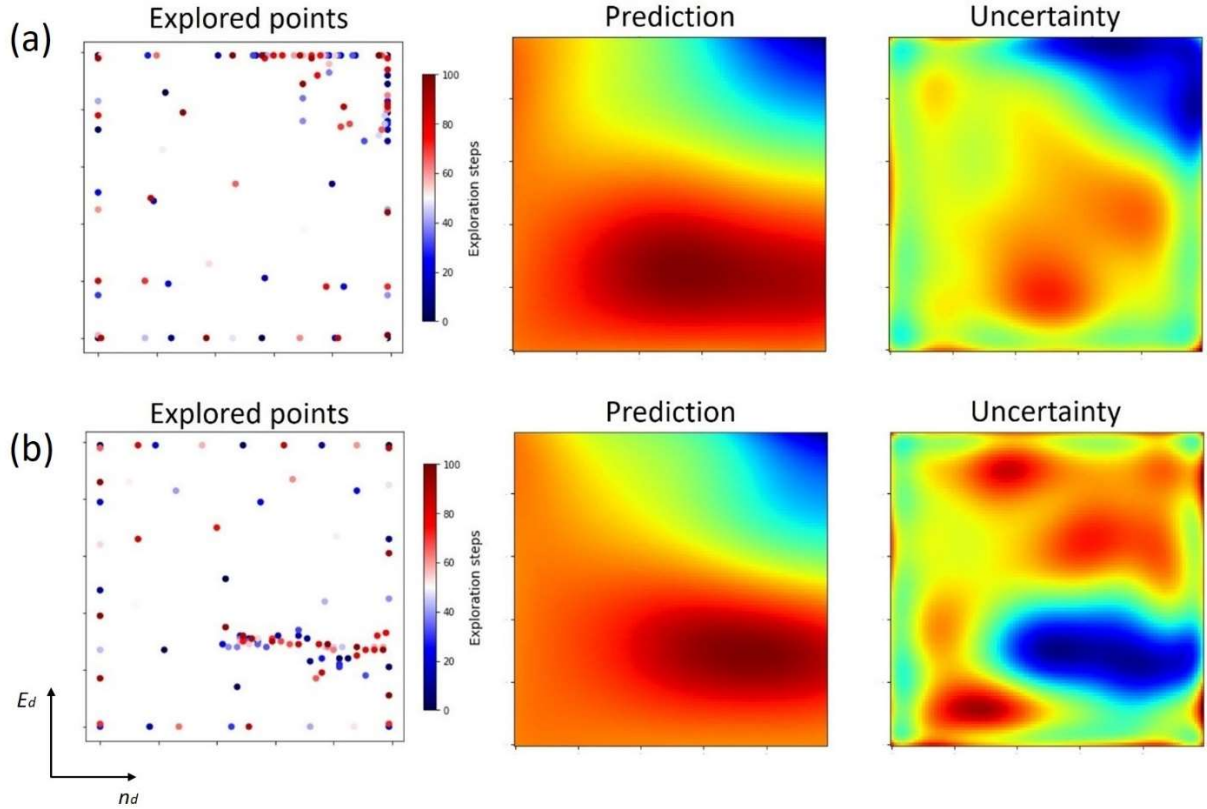

**Figure 6.** Bayesian optimization of the area under the loop as a function of random field and strength and defect concentration. Shown are exploration histories, prediction and uncertainty maps after 100 iteration steps for (a) target value of 0.7 and (b) 0.9.

This process is further illustrated in Fig. 6 showing the exploration histories, predictions, and uncertainty maps for the hysteresis loop area for different target values. Here, when the target value is chosen to be 0.7, the search points are concentrated in the top right corner of the parameter space, leading to the reduced uncertainty. The change of the target value to 0.9 concentrates the search in the vicinity of the maximum. Remarkably, the prediction maps converge very rapidly and are very close to each other.

The same optimization process can be extended to 3- and 4D space, including e.g. depolarization field, coupling strength, or field limits. However, visualization of the resulting dependencies in this case is more complicated, and hence is not shown.

**Conclusions**

To summarize, here we suggest and implement a Gaussian Process based methods to effectively explore and exploit the multidimensional parameter space of the complex non-local models to discover the regions of interest, in which the desired response adopts maximal value, i.e. yields desired functionalities. We discuss the specific adaptation of the acquisition function and sampling function to make the process efficient and balance the efficient exploration of parameter space for multiple possible minima and exploitation to densely sample the regions of interest where target behaviors are optimized.

This approach is illustrated via the hysteresis loop engineering in ferroelectric materials using continuous lattice model. However, it can be adapted in a straightforward manner to other functionalities and generative models, complementing previously developed stochastic optimization approaches.[47,48] The interactive notebook that allows reproducing the manuscript results is freely available at https://git.io/Jftuy.

**Data Availability:**

The data that support the findings of this study are openly available at [github.com/ramav87/Ferrosim].

**Acknowledgements:**

This effort (Gaussian Process) is based upon work supported by the U.S. Department of Energy (DOE), Office of Science, Basic Energy Sciences (BES), Materials Sciences and Engineering Division (S.V.K., R.K.V.) and was performed and partially supported (M.Z.) at the Oak Ridge National Laboratory's Center for Nanophase Materials Sciences (CNMS), a U.S. Department of Energy, Office of Science User Facility.